\documentclass[pre,twocolumn]{revtex4}
\usepackage[english]{babel}
\usepackage{amsmath}
\usepackage{amssymb}
\usepackage{epsfig}

\begin{document}
\title{Capillary Flotation in a System of Two Immiscible Bose-Einstein Condensates}
\author{Victor P. Ruban}
\email{ruban@itp.ac.ru}
\affiliation{Landau Institute for Theoretical Physics RAS,\\
Chernogolovka, Moscow region, 142432 Russia}

\date{\today}

\begin{abstract}
A spatially inhomogeneous, trapped two-component 
Bose-Einstein condensate of cold atoms in the phase
separation mode has been numerically simulated. 
It has been demonstrated for the first time that the surface
tension between the components makes possible the 
existence of drops of a denser phase floating on the 
surface of a less dense phase. Depending on the harmonic 
trap anisotropy and other system parameters, a stable
equilibrium of the drop is achieved either at the poles 
or at the equator. The drop flotation sometimes persists
even in the presence of an attached quantized vortex.
\end{abstract}

\maketitle

\section{Introduction}

Multicomponent mixtures of ultracold Bose-Einstein-condensed 
atomic gases have been studied for twenty-five
years [1-5]. Such systems consist either of different
chemical (alkaline) elements or of different isotopes of
the same element, or of the same isotopes in two 
different internal (hyperfine) quantum states. Interactions 
between components lead to a variety of phenomena 
that are absent in simple Bose-Einstein condensates. 
In this case, it is essential that the parameters
of nonlinear interactions between matter waves, 
proportional to the corresponding scattering lengths, 
in many cases can be changed over a wide range using
Feshbach resonances [6-10]. In particular, a sufficiently 
strong cross-repulsion between the two types of atoms 
results in spatial separation of condensates [11, 12] 
and in the presence of an effective surface tension
on the domain walls between the phases [4, 13]. This
separation is responsible for many interesting 
configurations and phenomena such as a nontrivial geometry
of the ground state of binary immiscible Bose-Einstein 
condensates in traps [14-16] (including optical
lattices [17-19]), bubble dynamics [20], quantum
analogs of classical hydrodynamic instabilities 
(Kelvin–Helmholtz [21, 22], Rayleigh-Taylor [23-25],
Plateau–Rayleigh [26]), parametric instability of 
capillary waves at the interface [27, 28], complex textures
in rotating binary condensates [29-31], core filled
vortices [3, 32-37], three-dimensional topological
structures [38–43], etc.

Large-scale dynamics of the interface in a separated 
binary condensate is very similar to the dynamics
of bubbles and drops in the classical mechanics of
immiscible ideal fluids [20, 23-25]. In the absence of
quantized vortices, the flow is potential inside each of
the components, and the entire vorticity of the velocity
field is concentrated at the interface. In this sense, the
bubble boundary is a vortex sheet similar to vortex
sheets in $^3$He-$A$ in some cases [44]. Such structures
also play an important role in turbulence (see [45-47]
and references therein).

Note that the analogy with classical hydrodynamics 
has not yet been exhausted. In particular, in the
presence of a sufficiently strong surface tension, a
moderate drop of a denser fluid can float on the surface 
of a less dense fluid. As far as I know, such capillary 
flotation in the case of binary condensates has not
yet been investigated. The aim of this work is to
numerically demonstrate the possibility of a heavy
drop to float within the framework of the system of
coupled Gross-Pitaevskii equations (1) and (2)
describing a dilute Bose-Einstein condensate at zero
temperature. Contrary to ordinary nearly incompressible 
fluids in a homogeneous gravitational field, we
will consider Bose-Einstein condensates in a trap
with a quadratic anisotropic potential, so that the
equilibrium profile of a less dense fluid $\rho_{\rm{eq}}({\bf r})$ 
[see Eq.(6)] will be strongly inhomogeneous in space. In this
case, the surface tension is proportional to $\rho_{\rm{eq}}^{3/2}$ [13];
i.e., it is absent on the Thomas-Fermi surface and
increases deeper into the condensate. Moreover, since
the Thomas-Fermi surface has the form of an oblate
or prolate ellipsoid of revolution, the effective potential 
energy of a drop depends not only on the locally
vertical deviation of its center of mass but also on two
local horizontal coordinates on the ellipsoid. As a
result, the drop can oscillate along the ``latitudinal''
coordinate of the ellipsoid during its swimming.
Depending on the system parameters, the horizontal
potential energy minimum can be achieved either at
the poles or at the equator of the ellipsoid.

\section{Model and numerical method}

The essence of the phenomenon can be demonstrated 
for the simplest case with equal masses of both
types of atoms $m_1=m_2=m$, which approximately
includes the case of a small difference in isotope
masses, for example, $^{85}$Rb and $^{87}$Rb. Let an 
axisymmetric harmonic trap be characterized by a transverse
frequency $\omega_\perp$ and anisotropy 
$\lambda=\omega_\parallel/\omega_\perp$. Choosing
scales $\tau=1/\omega_\perp$ for time, 
$l_{\rm tr}=\sqrt{\hbar/\omega_\perp m}$ for length, and
$\varepsilon=\hbar\omega_\perp$ for energy, dimensionless 
equations of motion for the complex wavefunctions $A({\bf r},t)$ 
and $B({\bf r},t)$ can be written in the form
\begin{eqnarray}
i\dot A=-\frac{1}{2}\nabla^2 A+
\left[V(x,y,z)+g_{11}|A|^2+g_{12}|B|^2\right]A,&&
\label{GP1}\\
i\dot B=-\frac{1}{2}\nabla^2 B+
\left[V(x,y,z)+g_{21}|A|^2+g_{22}|B|^2\right]B,&&
\label{GP2}
\end{eqnarray}
where $V=(x^2+y^2+\lambda^2 z^2)/2$ is the potential of the
trap and $g_{\alpha\beta}$ is the symmetric matrix of nonlinear
interactions (in our case, with positive elements).
Physical interactions are determined by scattering
lengths $a_{\alpha\beta}$ [2]:
\begin{equation}
g^{\rm phys}_{\alpha\beta}=2\pi \hbar^2 a_{\alpha\beta}
(m_\alpha^{-1}+m_\beta^{-1}).
\end{equation} 
Without loss of generality, the first self-repulsion coefficient can 
be set to $g_{11}=1$, since $g_{\alpha\beta}$ are considered 
to be fixed parameters (in each numerical experiment) in this work.

The conserved numbers of trapped atoms are given by the formulas
\begin{eqnarray}
&&N_1=\frac{l_{\rm tr}}{4\pi a_{11}}\int |A|^2 d^3{\bf r}
=(l_{\rm tr}/a_{11}) n_1,\\
&&N_2=\frac{l_{\rm tr}}{4\pi a_{11}}\int |B|^2 d^3{\bf r}
=(l_{\rm tr}/a_{11}) n_2.
\end{eqnarray} 
In real experiments, the ratio $l_{\rm tr}/a_{11}$ is in the range
from several hundred to several thousand.

The simple model specified by Eqs. (1) and (2) is
conservative. It is applicable only in the zero-temperature 
limit and cannot describe any finite-temperature
effects (including dissipative effects). For comparison,
the equations of motion for, e.g., $^3$He [48] are more
complex, and thermodynamics is of great importance.

The system of Eqs. (1) and (2) describes in essence
a two-fluid ideal hydrodynamics (when both velocity
fields are potential), with the difference that the total
pressure of each component consists of the normal
hydrodynamic pressure and the so-called quantum
pressure. Quantum pressure only appears where the density 
changes drastically in space. In our case, this is a
transition layer. The hydrodynamic pressures of both
fluids are $g_{11}|A|^4/2$ and $g_{22}|B|^4/2$, so that, 
at equal pressures, the component with the lower self-repulsion
coefficient is denser. For definiteness, it is assumed
that $g_{22}<g_{11}=1$; i.e., a relatively small drop of the
denser second phase will float on the free surface of
the first phase owing to the surface tension.

The background density profile of the first component 
is characterized by the chemical potential $\mu$. For
$\mu\gg 1$, in the Thomas-Fermi approximation, one obtains
\begin{equation}
|A_0|^2\approx [\mu-V(x,y,z)]\equiv\mu \rho_{\rm{eq}}({\bf r}).
\label{AA_eq}
\end{equation}
Thus, the effective transverse radius of the condensate is 
$R=\sqrt{2\mu}$, and the longitudinal half-size is $Z=\lambda R$.

The phase separation condition has the form 
$g=(g^2_{12}-g_{11}g_{22})>0$ [11, 12]. There is a narrow 
transition layer between separated condensates, where the
densities of both phases decrease to almost zero in one or
the other direction. In this case, the corresponding excess 
of energy (surface tension coefficient) is determined by a formula 
$\sigma=F(g_{22}/g_{11},g_{12}/g_{11}) |A_0|^{3}$ [11, 13]. 
In contrast to ordinary incompressible fluids, the
dependence of the surface tension on the background
density is very significant here.

The numerical modeling carried out in this work was focused on 
experimentally implemented mixtures $^{85}$Rb-$^{87}$Rb [8], where
$a_{12}/a_{22}\approx 2$, while $a_{11}$ can be changed over 
a wide range using the Feshbach resonance. Therefore, in all our 
numerical experiments, $g_{12}=2g_{22}$. There were different 
values for $g_{22}$, including $g_{22}=0.6,0.7,0.8$.

The used numerical method was quite standard
and included two procedures. The first procedure is
the preparation of the initial state by applying (to an
arbitrary ``input'' state) gradient descent, which in this
case is equivalent to imaginary time propagation on a
finite interval of an auxiliary quasi-time variable. This
dissipative procedure filtered out fast excitations and left
only the soft modes that did not reach their minimum
energy. A quasi-equilibrium transition layer was also
formed between the phases. The second procedure is
the actual modeling of the system of Eqs. (1) and (2)
using the second-order split-step Fourier method with
periodic boundary conditions in spatial coordinates.
In general, the numerical method is similar to that used in [49].

\section{Results}

\begin{figure}
\begin{center}
\epsfig{file=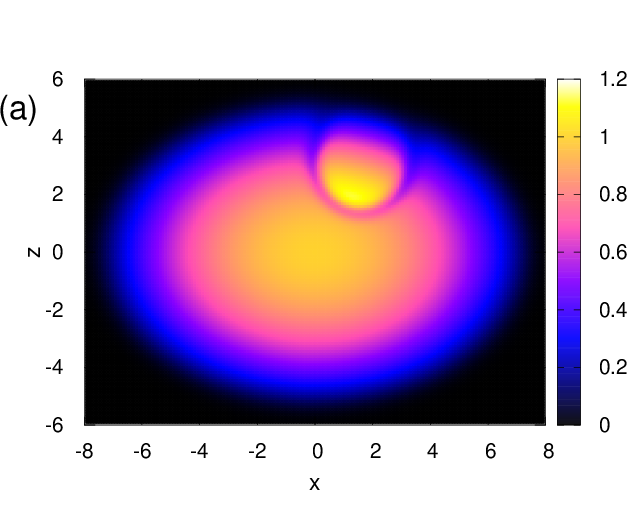, width=50mm}\\
\epsfig{file=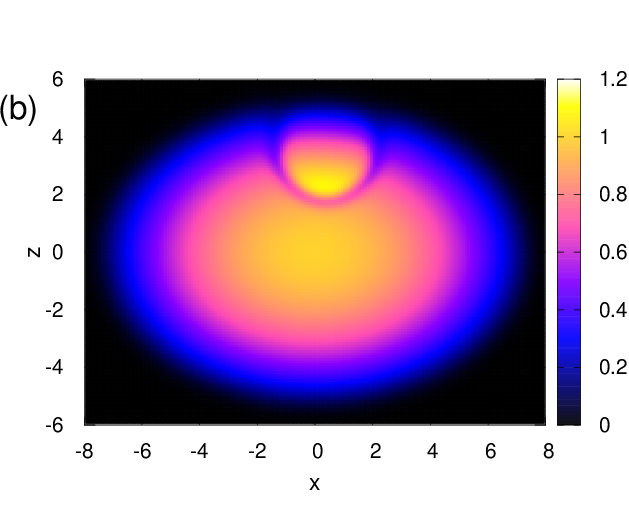, width=50mm}\\
\epsfig{file=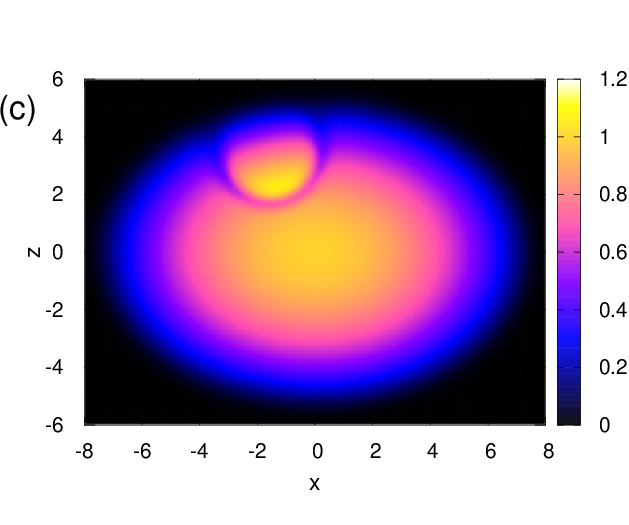, width=50mm}
\end{center}
\caption{
Total density map of two condensates normalized to $\mu$ in the $y=0$ 
plane for a floating heavy drop obtained numerically with the parameters
$\lambda=1.4$, $g_{11}=1.0$, $g_{22}=0.7$, $g_{12}=1.4$, $n_1=1282.4$, 
$n_2=37.1$, and $\mu=30$ for the times $t =$ (a) 500, (b) 600, and (c) 700. 
The drop undergoes slow oscillations near the pole, and panels (a) and (c)
correspond approximately to its extreme positions.
}
\label{example1} 
\end{figure}

\begin{figure}
\begin{center}
\epsfig{file=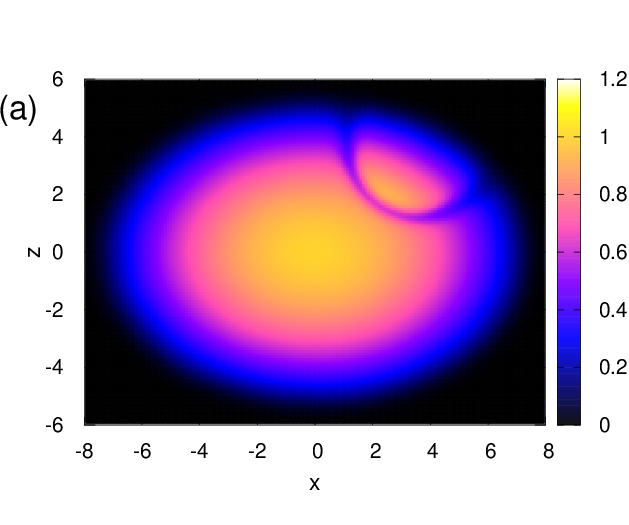, width=50mm}\\
\epsfig{file=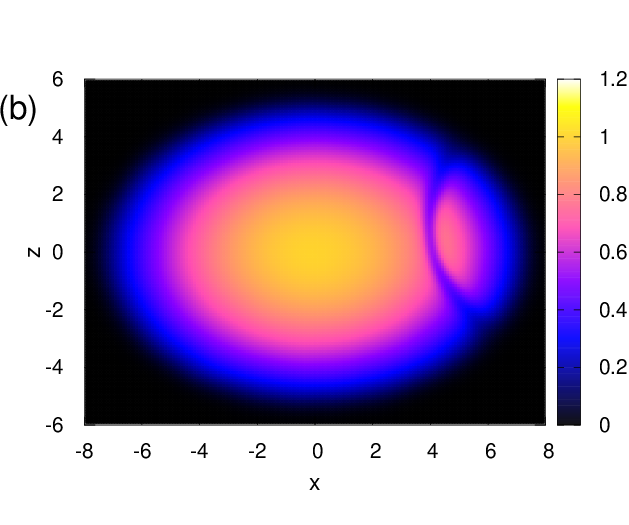, width=50mm}\\
\epsfig{file=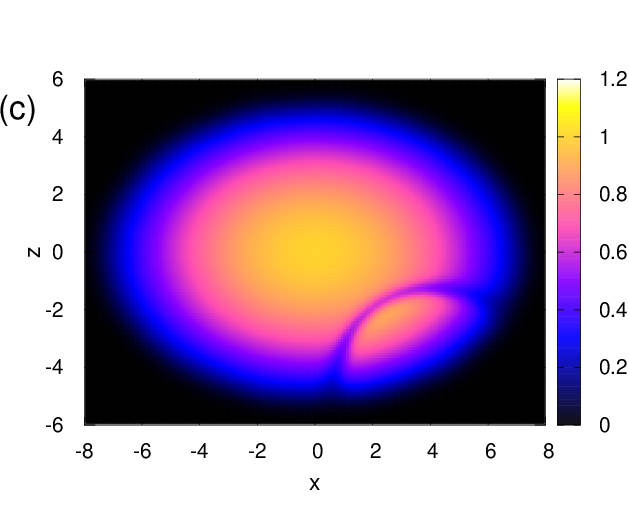, width=50mm}
\end{center}
\caption{
Total density map of two condensates normalized to $\mu$ in the $y=0$
plane for a floating drop oscillating near the equator. The drop is less 
dense, larger, and more mobile than that in Fig.1. It is obtained 
numerically with the parameters
$\lambda=1.4$, $g_{11}=1.0$, $g_{22}=0.8$, $g_{12}=1.6$, 
$n_1=1270.8$, $n_2=50.2$, and $\mu=30$
for the times $t =$ (a) 680, (b) 700, and (c) 720. Panels (a) and (c) 
correspond approximately to the extreme positions of the drop.
}
\label{example2} 
\end{figure}

\begin{figure}
\begin{center}
\epsfig{file=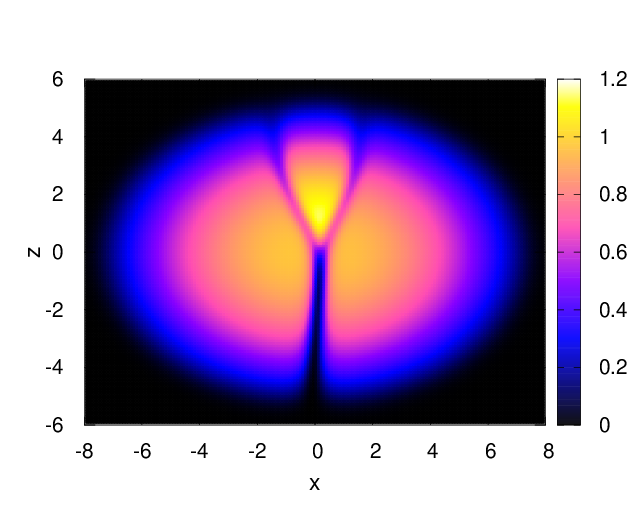, width=50mm}
\end{center}
\caption{
Total density map of a floating drop with an attached quantized vortex
in the $y = 0$ plane obtained numerically with the parameters 
$\lambda=1.4$, $g_{11}=1.0$, $g_{22}=0.8$, $g_{12}=1.6$, 
$n_1=1275.5$, $n_2=28.9$, and $\mu=30$ for the time $t =$ 300.
}
\label{example3} 
\end{figure} 

A natural tendency toward an increase in the maximum possible 
mass of a floating drop with a decrease in the relative 
difference between $g_{11}$ and $g_{22}$ was observed.

Some of the numerical results are demonstrated in Figs.1-3. 
For clarity, in the first and second examples, 
the configurations are chosen symmetric with
respect to the $y=0$ plane. The total density profiles in
this plane indicate the presence of a relatively denser
drop near the surface as well as a transition layer
between the components.

Fig.1 illustrates the case where the poles are stable 
positions of the drop on the ellipsoid. It oscillates
slowly around its high point (see video [50]). It should
be mentioned that a drop with a greater mass ($n_2=45$)
is no longer able to adhere to the surface and instead
undergoes complex motions inside the ellipsoid, sometimes 
remaining near the surface for a short time (see video [51]).

Fig.2 corresponds to a smaller relative difference
in the densities of the two phases compared to the first
example. In this case, the effective potential energy of
a floating drop reaches a local minimum at the equator
(see video [52]). In this example, the drop is sufficiently 
large and is deformed noticeably during its motion. 
Note that, in the case of a prolate ellipsoid
with anisotropy $\lambda=0.7$, the poles are stable positions
of such a drop (not presented here). Apparently, the
difference in behavior is due to different natures of the
dependences of curvature of the Thomas-Fermi surface 
and the local ``gravity'' on the polar angle. This is however
a problematic question, since for, e.g., $g_{22}=0.7$ (the
same as in Fig.1) the prolate ellipsoid still attracts the
drop to the pole (not shown). More research is required here.

It is interesting that a drop can also remain floating
when a quantized vortex passing through the first
component is attached to it (compare with [43], where
a large dense drop is located in the center of the system
and stabilizes several attached vortex filaments by its
mass). In the example shown in Fig.3, the drop does
not sink, although the vortex creates an additional
force pulling it down (see video [53]). A more massive
drop with $n_2=49.0$ under similar conditions moves
up and down repeatedly along the vortex from pole to
pole through the entire condensate (see video [54]).

\section{Conclusions}

To summarize, analogous to ordinary drops floating 
on the surface of a less dense fluid in a gravity field
owing to the surface tension, floating quantum drops
in a system of two immiscible trapped Bose-Einstein
condensates have been numerically detected. An
essential difference is the absence of a neutral 
equilibrium of a quantum drop in coordinates along the
Thomas-Fermi surface. This effect is obviously due to
the finiteness of the drop size compared to the size of
the main phase. An approximate analytical calculation
of the corresponding dependence of the drop potential
energy and its equilibrium geometric shape remains an
interesting problem for further investigation.

\end{document}